%% file: main.tex
\documentclass[11pt]{article}

\usepackage{arxiv}
\usepackage[utf8]{inputenc}
\usepackage[T1]{fontenc}
\usepackage{lmodern}
\usepackage{microtype}
\usepackage{graphicx}
\usepackage{amsmath,amssymb,amsfonts}
\usepackage{booktabs,longtable,array}
\usepackage{hyperref}
\usepackage{natbib}
\usepackage{xcolor}
\usepackage[section]{placeins}
\usepackage{tikz}
\usetikzlibrary{arrows.meta,positioning}
\graphicspath{{figures/}}
\hypersetup{colorlinks=true,linkcolor=blue!50!black,citecolor=blue!50!black,urlcolor=blue!50!black}
\renewcommand{\shorttitle}{Operationalising weaker faithfulness}
\renewcommand{\headeright}{}
\renewcommand{\undertitle}{}

\renewenvironment{abstract}
  {\centerline{\large\bfseries Abstract}\begin{quote}}
  {\end{quote}}

\title{Operationalising Weaker Faithfulness:\\
Recovery, Ambiguity and Computational Limits}

\author{
  Cillian Hourican\thanks{\raggedright Computational Science Lab, Informatics Institute, University of Amsterdam. Corresponding author: \texttt{c.j.hourican@uva.nl}.} \\
  \And
  Rick Quax\thanks{Computational Science Lab, Informatics Institute, University of Amsterdam; \texttt{qr.quax@uva.nl}.} \\
}
\date{14 October 2026}

\begin{document}
\raggedbottom
\maketitle

\begin{abstract}
\noindent Variables can predict an outcome jointly while each is individually
independent of it, causing structure learners to discard informative links.
We investigate how to operationalise weaker faithfulness through two approaches:
selective protection in an adjacency learner and explicit retention of
alternative directed graph explanations. Matched experiments show that
changes at both an early candidate-removal stage and later pruning can be
necessary. In noisy XOR with a descendant ($N=10{,}000$, 20 matched samples),
combining reactivation and protection raises complete parent-pair recovery
from $0/20$ for either change alone to $20/20$, while adding false edges.
A confirmation study shows that the family estimator can preserve
explanations lost by orienting the protected learner's selected skeleton:
in one weaker-signal XOR condition ($N=10{,}000$) it retains the entire reference
family in $48/50$ samples, compared with $0/50$ for the skeleton adapter.
These gains are conditional. Protection can add only error, weaker effects
produce false exclusions, and an ordinary faithful-chain comparison favours
the PC learner. Complete nonempty family outputs decline at six/eight variables, and
all 400 twelve/twenty-variable searches remain unfinished at the reporting
cutoff. Neural controls show engineered retention, while the family estimator
returns no accepted graph within its application budget. This operationalisation
identifies two deletion stages that
must be coordinated in the motivating example and shows when direct family
search preserves explanations lost by selecting one skeleton. Finite evidence
and computation remain substantial barriers to reliable general inference.
\end{abstract}

\input{Sections/1_introduction}
\input{Sections/2_methods}
\input{Sections/3_results}

\input{Sections/4_discussion}
\input{Sections/5_conclusion}

\section*{Code and data availability}
The code and derived data supporting this study are available from the
authors upon reasonable request. The neural recordings analysed in this
study are available from CRCNS \citep{ito2016ssc3}.

\section*{Funding}
This work was supported by the Netherlands Organisation for Health Research
and Development (ZonMw) under Open Competition Grant 09120012010063.

\bibliographystyle{plainnat}
\begingroup\small
\bibliography{references_phase3}
\endgroup
\clearpage
\appendix
\input{Sections/A_appendix}

\end{document}

%% file: Sections/1_introduction.tex
\section{Introduction}\label{sec:intro}

A relationship can be invisible when variables are examined separately and
clear when they are considered together. Suppose two independent fair binary
variables, $X$ and $Y$, determine a third through noisy exclusive OR (XOR):
$Z=X\oplus Y\oplus E$, where $\oplus$ denotes addition modulo two and the
independent error $E$ flips the output with probability
$0<\epsilon<1/2$. Neither parent alone predicts $Z$. Knowing both does.
A structure learner that permanently removes the links to $Z$ after testing
only their marginal associations loses the very relationship we want to find.

This example challenges a familiar assumption in causal discovery.
The causal Markov condition says that separations in a directed acyclic graph
(DAG) imply statistical independences. Faithfulness supplies the converse:
independences in the distribution arise from those graphical separations
\citep{spirtes2000causation}. Noisy XOR violates faithfulness to its generating
collider $X\to Z\leftarrow Y$, despite having strictly positive probabilities.
Here a collider is a pair of arrows meeting at a common target. The example
therefore concerns more than a deterministic boundary case or an unfortunate
finite sample.

Weaker-faithfulness assumptions allow some relationships to be supported by
triple interactions rather than by persistent pairwise dependence.
\citet{marx2021weaker} developed such assumptions and a corresponding
Markov-blanket learning procedure. A Markov blanket is a set that renders
a target independent of the remaining variables; in a DAG its graphical
form consists of parents, children and other parents of those children.
Making the theoretical idea operational requires choices about candidate
search, deletion, finite evidence and the meaning of the returned graph.
These choices determine whether a jointly informative relationship survives
and what can subsequently be inferred from it.

Our purpose is to investigate that operationalisation. We study two concrete
approaches. The first modifies an adjacency learner: an adjacency is a link
without an assigned direction, and the set of these links is the graph's
\emph{skeleton}. We allow a candidate to survive an early deletion decision
when further conditioning restores dependence, and selectively protect
surviving links from later pruning using a joint-information contrast.
The second approach searches for a \emph{family of DAGs}: alternative directed
structures satisfying an explicitly specified set of weaker-faithfulness assumptions.
These procedures have different targets. The protected learner selects one
skeleton; the family estimator attempts to retain alternative explanations,
including different skeletons.

The distinction matters even in the three-variable example. Under the
weaker adjacency assumption combined with subgraph minimality, any of the
three variables can occupy the collider role. These three alternatives were
already identified by \citet[Section~7.1]{marx2021weaker}. Selecting one
skeleton and orienting it cannot restore the other two. A more recent
approach, MD-PC, also represents candidate graph classes under its own
minimal-dependence assumptions \citep{ramazi2026minimal}. We identify two
places where a joint relationship can be deleted, test interventions at both,
and evaluate whether direct family search preserves explanations lost by
selecting one skeleton.

The experiments address three connected questions. Which deletion stages
must change to retain a joint relationship? What is gained, and what is
lost, when alternative graph explanations are kept explicitly? How far do
those gains survive finite sampling and increasing problem size? We combine
matched stage comparisons, exact small-law reference calculations, a
confirmation study and an application on a fixed neural background.
The exact construction provides a reference; its finite implementation is
an approximation whose reliability is evaluated separately.

The results establish useful but conditional gains. Coordinated changes
recover parent links that either change alone can lose. Family search
retains explanations excluded by a selected skeleton, including a signal
below the protection rule's threshold. These benefits coexist with extra
edges, false exclusions under weak evidence and severe limits on larger
searches. Static experiments carry the central recovery claims. Temporal
experiments distinguish independent episodes, supplied time order and
serial observations; the neural application tests operational reach rather
than establishing biological causal truth. Together, the results show what
this attempt to operationalise weaker faithfulness achieves and which
obstacles remain for a generally useful learner.

%% file: Sections/2_methods.tex
\section{Methods}\label{sec:methods}

\subsection{Retaining joint relationships through an adjacency search}\label{sec:search}

The first approach separates candidate admission from two opportunities to
remove candidates. Grow admits potentially related variables, shrink revisits
each target's candidate set, and global pruning tests the assembled links.
We change shrink and pruning independently to determine where a relationship
is lost. The resulting procedure is a bounded implementation motivated by
weaker faithfulness; it differs from the algorithm covered by the original
blanket-recovery theorem.

For each target $Z$, grow considers other variables in the fixed column
order. It admits a candidate when any tested conditioning set establishes
dependence, trying the empty set, then singletons and pairs from the remaining
variables. Original shrink removes a candidate if a nonempty singleton or
pair from the current candidate blanket yields independence from the target.
\emph{Reactivation shrink} examines the first such conditioning set, called
a separator. If it is a
singleton, adding one other current blanket member can restore dependence
and veto the removal. A separator already of size two is not extended.
The term reactivation thus denotes reconsideration of a proposed removal,
not unrestricted recovery of previously discarded candidates.

The retained candidate blankets determine an initial edge set $E_0$.
Static edges require reciprocal membership. Global pruning tests each edge
for independence given the empty set, singletons and pairs drawn from all
other variables, and removes it if a separator is found. Its conditioning
pool is fixed while testing the initial edge list. The adjacency learner's temporal
branch has different eligibility rules, stated in
Appendix~\ref{app:implementation}.

\emph{Protection} changes this last deletion stage. For each pair
$\{X,Y\}$ in the final candidate blanket of $Z$, calculate the empirical
information contrast, in bits,
\begin{equation}
 \widehat J(X,Y;Z)=\widehat I(X,Y;Z)-\widehat I(X;Z)-\widehat I(Y;Z).
 \label{eq:statistic}
\end{equation}
Mutual information $I$ measures information shared between variables;
$J$ compares joint information about $Z$ with the sum of the two individual
contributions. If $\widehat J>0.3$, the identifiers $\{X,Z\}$ and
$\{Y,Z\}$ are protected from pruning wherever they occur in $E_0$.
The rule cannot create an edge absent from $E_0$.

The fixed $0.3$-bit threshold is an effect-size choice; it supplies neither
a significance level nor a causal certificate. All pairs in the candidate
blanket are considered, including
pairs that are not causal co-parents.

\begin{table}[t]
\centering
\caption{Matched controls for protected adjacency learning. All arms use the
same grow procedure and conditional-independence test (nominal level $.05$, conditioning cap two,
no multiple-testing filtering). The stages are grow, shrink, edge assembly
and global pruning; protection affects only the last stage. R is the
no-pruning reference, outside the A--D factorial.}\label{tab:arms}
\begin{tabular}{llll}
\toprule
Arm & Shrink & Protection & Global pruning\\
\midrule
A & Original & Off & On\\
B & Reactivation & Off & On\\
C & Original & On & On\\
D & Reactivation & On & On\\
R & Reactivation & Off & Off\\
\bottomrule
\end{tabular}
\end{table}

The A--D factorial in Table~\ref{tab:arms} distinguishes stage effects.
D--B isolates protection after reactivation; C--A isolates protection after
original shrink. D--R compares selective protected pruning with no pruning.
It does not isolate a shrink change. Protection guarantees survival only for
protected edges already present at pruning; Appendix~\ref{app:implementation}
gives the corresponding set identity.

These arms use a modified likelihood-ratio $G^2$ test, using
a chi-square reference and nominal level $.05$. A nonrejection is used as an
independence decision. The conditioning cap and adaptive decisions do not
provide graph-level error control. The table-cell conventions are stated in
Appendix~\ref{app:implementation}.

\subsection{From a selected skeleton to alternative graph explanations}\label{sec:family}

The second approach starts with an explicit population target. Let $P$ be
a strictly positive joint distribution of finite categorical variables $V$,
and let $K$ denote independently supplied restrictions, such as a known
time order. Define
\begin{equation}
 \mathcal F^*(P,K)=\{G:\ G\text{ is a DAG on }V,\quad
 M(G,P),\ A_2(G,P),\ S(G,P),\ K(G)\}.
 \label{eq:family}
\end{equation}
Here $M$ is Markov compatibility: every graphical separation in $G$ implies
the corresponding independence in $P$. The weaker adjacency condition
$A_2$ allows each adjacent variable to be supported either individually or
through an appropriate triple association. Its quantifiers range over all
relevant conditioning sets; the ``two'' bounds the association group, not
conditioning depth. $S$ is subgraph minimality (also called SGS-minimality):
no proper directed subgraph of
$G$ is still Markov-compatible with $P$. It does not require the fewest
edges among all unrelated DAGs. Appendix~\ref{app:theory} specifies these
predicates, including the consequential strict-association convention.

A causal reading additionally requires an acyclic, causally sufficient
model without selection effects, with the generating graph satisfying the
specified assumptions. Enumeration does not establish these premises for
an observed system. The principal target does not add the source's
orientation-faithfulness assumption; that is a separate sensitivity.

\begin{figure}[htbp]
\centering
\input{figures/revision_xor_families}
\caption{Alternative explanations for independent fair-parent noisy XOR,
under Markov compatibility, the declared weaker adjacency condition and
subgraph minimality. The three DAGs have different missing adjacencies and
collider roles. Orienting one skeleton cannot recover all three, and their
edge union loses the joint constraints. This ambiguity is described by
\citet[Section~7.1]{marx2021weaker}. Time order supplies additional knowledge.}
\label{fig:xor-family}
\end{figure}

For small known laws we calculate $\mathcal F^*$ by exhaustive DAG filtering
with exact rational independence checks. A second construction uses each
ordering of the variables: an earlier variable is a parent of a later one
exactly when they remain dependent conditional on all other predecessors
of the later variable. For positive laws this yields the unique minimal
Markov DAG for that order. This established construction is discussed by
\citet{lam2022greedy}, building on earlier permutation methods. Across all
orders it supplies every subgraph-minimal DAG; applying the remaining predicates
returns the target in Equation~\ref{eq:family}. We retain all qualifying
candidates rather than selecting a globally sparsest graph.

We implement a practical version of this search with repeated-state merging,
cached queries and persistent storage. It uses finite
conditional-independence (CI) decisions instead of an exact law.
The selected test compares a summed $G^2$ statistic against 199 random tables
per informative conditioning stratum, sampled with the observed margins
fixed. A Monte Carlo $p$-value with the plus-one correction is rejected only
when $p<.05$. Nonrejections are used as independence decisions; unavailable
queries are never converted to independence. The policy was selected in a
separate calibration study before the confirmation data were generated.
That calibration concerns test behaviour under independent sampling, not
simultaneous graph coverage.

We call this implementation the \emph{weaker-faithfulness (WF) family estimator}.
Unless a different CI policy is specified, WF denotes this selected-policy
implementation throughout. Completion means
its declared traversal has finished. It does not imply that its output equals
the family obtained by applying every predicate to a complete table of
finite CI decisions: noisy decisions need not obey the probability identities
used by the exact-law construction. We evaluate that discrepancy explicitly.
Appendix~\ref{app:finite} defines the separate empirical reference and a
broader Markov confidence family used for uncertainty diagnostics.

For comparison, we enumerate compatible acyclic orientations of D's selected
skeleton, calling this output the \emph{D adapter}. It adds no CI evidence. Unresolved
triples remain unresolved; protection is not treated as collider evidence.
On at most four variables it represents the complete resulting family.
It therefore asks what can be recovered after committing to D's skeleton,
not how accurately a second algorithm estimates the same population target.
Above four variables the adapter seeks at most one accepted candidate DAG,
or \emph{witness}, and
retains an explicit incomplete status unless enumeration actually exhausts.

\begin{table}[htbp]
\centering\small
\caption{The reference families and estimated outputs answer different questions.}
\label{tab:family-targets}
\begin{tabular}{p{.25\textwidth}p{.68\textwidth}}
\toprule
Object & Meaning\\
\midrule
Population reference $\mathcal F^*$ & All DAGs satisfying the stated restrictions
for the known distribution.\\[3pt]
Empirical reference $\mathcal E$ & All DAGs satisfying those restrictions when
evaluated against the complete table of finite CI decisions.\\[3pt]
WF output & The result of the finite order search; even a completed
output can differ from $\mathcal E$.\\[3pt]
Markov confidence family $\mathcal C$ & A broader uncertainty construction
without the weaker adjacency or minimality restrictions.\\[3pt]
D adapter & Compatible orientations of the single skeleton selected by D.\\
\bottomrule
\end{tabular}
\end{table}

A family is represented by its candidate DAGs and any necessary joint
constraints. A display of all possible edges alone can admit combinations
that never occur in a candidate. We compress families into graphical
classes only when their complete extensions equal the retained candidates.
A complete empty WF family means the completed traversal retained no candidate;
an empty graph is instead one candidate with no edges. An unfinished search
can supply witnesses but cannot establish global absences or compelled
arrows---directions shared by every candidate in the complete family---from
those witnesses alone.

\subsection{Experiments and comparison targets}\label{sec:design}

We use $N$ for the number of rows per input; replicate counts refer to separate
input datasets. The adjacency experiments comprise a matched stage study and
a controlled expansion. The stage study used 120 four-variable inputs shared
across five arms and static/time-labelled modes. The controlled expansion used
35 conditions, two sample sizes ($2{,}000$ and $10{,}000$), and 20 replicates
per condition and size. It varied output noise, marginal signals, parent
probabilities and independent nuisance variables. The descendant baseline
sets $Z=X\oplus Y\oplus E$ with fair independent parents and flip probability
$.05$, and $W=Z\oplus F$ with independent flip probability $.1$.
The factorial evidence and its selective-pruning comparison motivate the
confirmation study; each design is reported with its own denominators.

The confirmation study comprises 6,400 new inputs across 148 cells.
Core cells use 50 independent inputs; scaling cells use 25. Sample sizes
are 500, 2,000 and 10,000 where specified in the design; scaling uses 2,000 and 10,000
with four, six, eight, twelve or twenty variables. The design includes
joint-effect and ordinary mechanisms, weak signals, known-assumption stress
cases, independent two-slice episodes, trajectories and matched independent
controls. Appendix~\ref{app:confirmation-design} defines the mechanisms behind
the reported examples. All conditions and exclusions remain in the supplementary
tables. Methods
share each input and the declared query-seed scheme; background-knowledge
views analyse the same input under different supplied restrictions and do
not count as new independent samples.

Confirmation includes B, D and R, the D adapter and the WF family estimator.
WF with the depth-two $G^2$ policy and exact empirical/uncertainty references diagnose
the role of finite evidence. PC-stable (PC below) is evaluated only in independently
classified faithful four-variable conditions; its stable skeleton search
\citep{colombo2014order} is followed by checked class representation.
Only the skeleton stage has the order-independence claim.
Appendix~\ref{app:comparators} describes these implementations and the
additional MD-PC comparisons under that method's own assumptions.

Population reference families are available for the supported small laws.
Larger experiments can evaluate membership of the planted DAG where
applicable, but do not have a full exact population-family reference.
The above-four-variable scaling laws also lack a verified classification
under the chosen weaker-faithfulness assumptions. Their failures are
practical outcomes, not counterexamples within an established theorem domain.
Deterministic boundaries and incompatible-assumption cases are labelled
separately from admitted positive laws.

\subsection{Time order and the neural application}\label{sec:application-design}

The adjacency experiments' temporal mode relabels static numeric columns;
it isolates an eligibility convention rather than simulating trajectories.
The confirmation distinguishes that convention from independent episodes
with an explicit baseline graph and from genuine overlapping trajectory
rows. A tier groups variables at the same supplied time level. Known tiers
forbid backward arrows but do not supply adjacencies,
baseline independence or a within-slice order. Matched independent controls
help describe sampling effects; they do not validate independent and identically
distributed (IID) tests for serial data.
When relevant history is omitted, accuracy against a causal graph for all
variables in both slices is unavailable even for the corresponding independent controls.

The application uses one CRCNS ssc-3 culture recording
\citep{ito2016ssc3,ito2014recording}. Binary occupancy in 5, 10 and 20\,ms bins
is paired with the next bin. The recorded selection contains 12,000 rows
per width, sampled without replacement in a fixed order. Selected transitions
can share an underlying bin; the CSV order is not chronological.
The 18-unit, 36-column adjacency application compares A--D/R and engineered
controls. Its observed graphs have no known causal ground truth.

The family application fixes the first six units from that selection
at both slices, giving twelve neural columns. Independent synthetic modules
add three or four columns: symmetric XOR, XOR with a descendant, ordinary OR,
and an independent negative control. Each profile/bin has 100 planned
synthetic randomizations conditional on the same neural background. D, its
adapter and WF are compared; descendant modules also have a
separate supplied-tier view. The full 36-column panel is unsupported by the
family implementation. Module-reference comparisons are local
benchmarks, not exact reference families for the combined neural distribution.

Each WF application attempt receives one 60-second elapsed search window.
D and WF have an external 90-CPU-second/120-wall-second attempt limit and
4\,GiB memory bound. The adapter has a 45-CPU-second/60-wall-second limit.
These limits define this operational demonstration; they differ from the
resumable confirmation searches. Appendix~\ref{app:application} records the
application endpoint and all-planned accounting.

\subsection{Outcomes, uncertainty and reporting boundaries}\label{sec:metrics}

Selected-skeleton recall and false-edge counts are measured against the
planted skeleton; complete parent-pair recovery requires both designated
parent links. For a nonempty population reference, \emph{whole-family
coverage} means every reference DAG is returned. \emph{Family equality}
means no reference DAG is lost and no extra DAG is added; it is also defined
when both families are empty. These differ from
membership of one planted DAG. An asserted edge, absence or direction is
false when it fails for at least one member of the relevant reference family.
False-assertion rates condition on complete nonempty outputs; we also report
availability and bounds over all planned applicable inputs. Empty and
unfinished outputs do not acquire assertions by vacuous intersection.
For example, each pair of variables is absent in one member of the
three-collider XOR reference (Figure~\ref{fig:xor-family}). Declaring any pair
definitely adjacent exceeds what this family identifies, even when that edge
occurs in the planted graph. This error differs from adding a false edge
relative to one planted skeleton.

For reporting, unfinished confirmation searches were discontinued at a cutoff
chosen after execution began. This was not a common prespecified cumulative
runtime limit. All planned inputs remain accounted for, including unfinished
searches. Main-text examples were selected
after analysis to explain the mechanisms and limits; full supplementary tables
retain all cells and sample sizes.

Confirmation binary intervals are two-sided 95\% Clopper--Pearson intervals;
continuous and paired intervals use the prespecified 10,000 whole-input
bootstrap draws. They are descriptive cellwise intervals without simultaneous
superiority interpretation. The adjacency-experiment figures retain their
Wilson/Monte Carlo uncertainty conventions, specified in captions.
Observed CPU and memory are reported with their measurement scope and
completion status; work observed before interruption is not eventual
completion time. Fixed neural-background randomizations do not provide
independent biological replication.

%% file: figures/revision_xor_families.tex
\begin{tikzpicture}[>=Stealth, every node/.style={font=\small},
  wfnode/.style={circle,draw,minimum size=7mm,inner sep=1pt}]
\foreach \wfoffset/\wfrole/\wfcaption in {0/Z/{Collider at $Z$},4.1/X/{Collider at $X$},8.2/Y/{Collider at $Y$}} {
  \begin{scope}[xshift=\wfoffset cm]
    \node[wfnode] (X\wfrole) at (0,0) {$X$};
    \node[wfnode] (Y\wfrole) at (1.4,0) {$Y$};
    \node[wfnode] (Z\wfrole) at (.7,-1.15) {$Z$};
    \node at (.7,.7) {\wfcaption};
  \end{scope}
}
\draw[->,thick] (XZ)--(ZZ); \draw[->,thick] (YZ)--(ZZ);
\draw[->,thick] (YX)--(XX); \draw[->,thick] (ZX)--(XX);
\draw[->,thick] (XY)--(YY); \draw[->,thick] (ZY)--(YY);
\node[align=center,text width=11cm,font=\small] at (4.8,-2.15)
 {The same joint distribution permits three collider roles.\\
 Knowing that $X,Y$ precede $Z$ retains only the left-hand graph.};
\end{tikzpicture}

%% file: Sections/3_results.tex
\section{Results}\label{sec:results}

\subsection{Coordinating deletion stages recovers relationships}\label{sec:stage-results}

\begin{figure}[!ht]
\centering
\includegraphics[width=\textwidth]{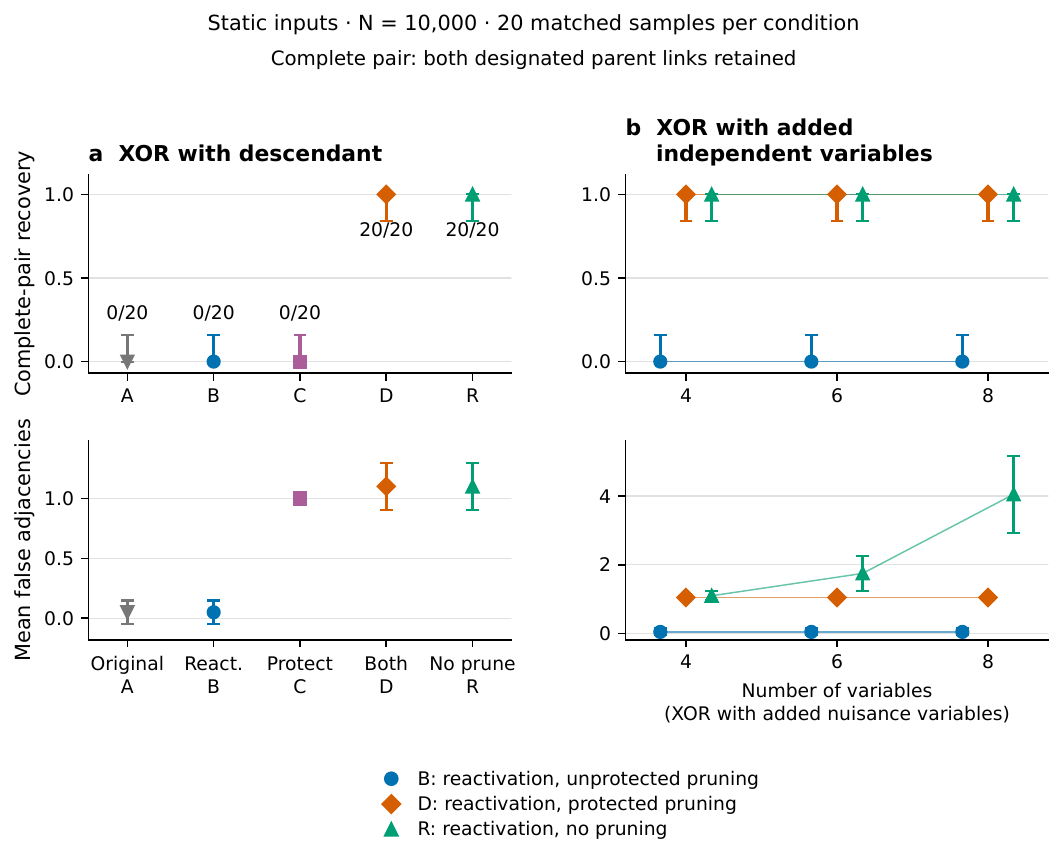}
\caption{Operational choices affect both retention and errors. The matched
stage comparison uses static noisy XOR with a descendant; the nuisance-variable
comparison uses XOR with added independent variables. Both panels draw on
the static controlled expansion, with 20 matched replicates per
condition at $N=10{,}000$. Complete parent-pair recovery requires both designated
parent links and differs from whole-skeleton recall. Uncertainty retains the
95\% Wilson intervals for binary
recovery and normal Monte Carlo intervals for mean false-edge counts.
The no-pruning reference R is interpreted separately from the A--D factorial.
The confirmation study uses separate samples.}
\label{fig:retention}
\end{figure}

The matched stage comparison explains why adding a protection rule is not
sufficient on its own. In static noisy XOR with a descendant at
$N=10{,}000$, the original learner (A), reactivation without protection (B)
and protection after original shrink (C) each recover a complete parent pair
in $0/20$ samples. Their combination (D) recovers the pair in $20/20$
(Figure~\ref{fig:retention}). The traces locate the two losses. In one saved
sample, original shrink removes $X$ from $Z$'s candidate set after conditioning
on descendant $W$ ($p=.947$). Adding $Y$ restores detected dependence and vetoes
that removal. Reactivation alone remains insufficient: unprotected pruning
subsequently removes $X-Z$ marginally ($p=.151$). With protection, the $.705$-bit
joint-information contrast preserves the link. The same sequence occurs for
$Y-Z$. Within A--D, both interventions are needed in this condition.

The gain has a surrounding-graph cost. In the same samples, adding protection
after reactivation lowers mean skeleton precision from $.975$ to $.738$ and
adds 1.05 false adjacencies on average. Every edge retained by the protection
rule still needs to be interpreted as a candidate relationship. The statistic
does not establish that the two protected links are causal parent links.
With independent nuisance variables, selective protection can retain the
pairs also recovered by disabling pruning while removing many unwanted
links. Thus the useful design question is which links to preserve, rather
than whether to abandon deletion altogether.

The confirmation study supports the retention mechanism and its conditional scope.
In symmetric XOR with flip probability $.10$ at $N=10{,}000$, protection
raises mean adjacency recall from $.02$ for B to $1.00$ for D; the paired
difference is $.98$ with descriptive 95\% interval $[.95,1.00]$ across
50 inputs. Mean false edges rise from $.02$ to $1.00$, a difference of
$.98\,[.94,1.00]$. This is recovery of previously missed adjacencies at a
false-edge cost, not correct-skeleton selection.

Protection sometimes offers no benefit or only a cost. With XOR flip
probability $.25$, both B and D have recall $.03$ and mean false edges $.06$.
In an asymmetric interaction with unequal parent probabilities ($.35$ and
$.60$) and unequal output-error probabilities
(Appendix~\ref{app:confirmation-design}), the paired recall gain is zero,
while protection adds $.94\,[.86,1.00]$ false edges. Conversely, ordinary OR
with $.10$ output flips gives recall $1.00$ for both D and the no-pruning
reference R, with mean false edges $.02$
and $1.06$, respectively. The same implementation can therefore recover a
hidden relationship, preserve an unnecessary edge, miss a weak signal or
usefully remove a nuisance link. The fixed rule has no uniform advantage.

\subsection{Keeping alternative explanations changes what can be recovered}\label{sec:family-results}

The exact noisy-XOR reference illustrates why adjacency retention and causal
identification are different tasks (Figure~\ref{fig:xor-family}). With no
ordering knowledge, its three collider explanations have different skeletons.
Their union is a triangle, which is not itself any of the three explanations.
A learner may preserve both planted parent links yet commit to a skeleton
that cannot represent the full answer. An independent fourth variable does
not change the three alternatives.

The finite comparisons show that this distinction can matter in useful
signal regimes. In the strong XOR confirmation cell (flip $.10$,
$N=10{,}000$), WF covers the entire reference family in $42/50$ samples,
compared with $0/50$ for the D adapter; the paired difference is
$.84\,[.74,.94]$. WF equals the reference in $38/50$ and returns a complete
empty family in seven samples. There are no evaluable inputs for orientation conditional on a correctly
selected D skeleton. Its failed reference coverage consequently cannot
be presented as poor orientation conditional on a correct skeleton.

In the weaker flip-$.45$ XOR cell, WF covers the reference in $48/50$ samples
and equals it in $41/50$, with mean family size $3.26$. The D adapter again
has coverage $0/50$; D and B each have mean adjacency recall $.01$ and no
complete parent-pair recovery. This law's joint-information contrast lies
below the fixed $.3$-bit protection threshold. The family approach thus has
useful reach beyond that retention rule. The comparison changes the upstream
search and CI evidence together; it does not isolate the contribution of one
component.

\begin{figure}[htbp]
\centering
\includegraphics[width=\textwidth]{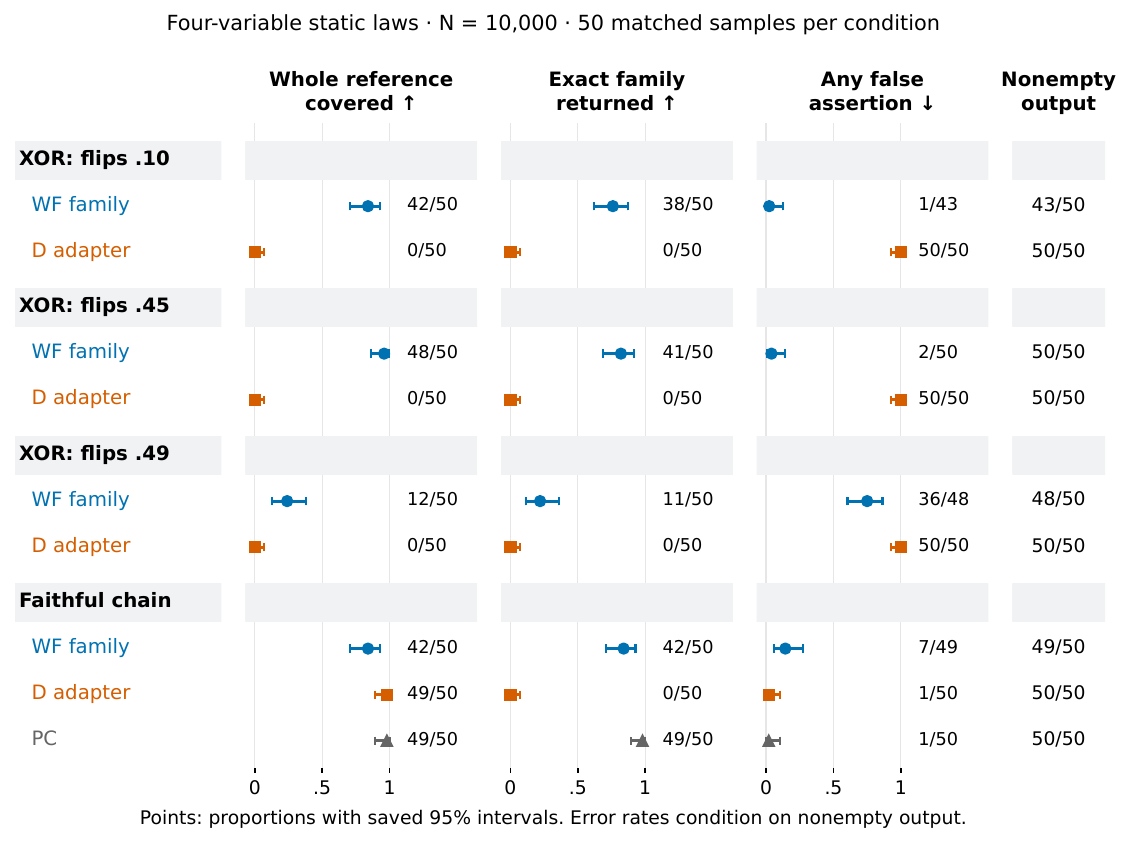}
\caption{Recovery, error and availability in four explanatory confirmation
cells, all four-variable static inputs with $N=10{,}000$ and 50 independent
samples. All methods are compared with the same population WF reference in
each cell. Coverage requires every reference DAG; equality also excludes extra
DAGs. False assertions contradict at least one reference member and are scored
only for complete nonempty outputs, with those denominators shown. Points and
intervals retain saved 95\% cellwise estimates. PC is included
only in its eligible faithful-chain cell. The D adapter conditions on D's
selected skeleton and adds no CI evidence. Cells were selected after analysis
for exposition; all planned cells remain in the full supplementary confirmation
tables.}
\label{fig:family-tradeoffs}
\end{figure}

Coverage drops sharply in the near-independent flip-$.49$ XOR cell
(Figure~\ref{fig:family-tradeoffs}): WF coverage is $12/50$ and equality $11/50$; at
$N=500$ and $2{,}000$, coverage is only $0/50$ and $1/50$. At $N=10{,}000$,
$36/48$ nonempty outputs make a false structural assertion. These are
principally false \emph{absence} claims: the mean number of incorrectly
excluded adjacencies is $2.23$ among those 48 outputs, while the mean false
directed-arc count is zero.
The all-planned false-assertion bounds are $[.72,.76]$.

In a faithful four-node chain, where each node copies its predecessor with
$.10$ flip probability, WF covers the reference in $42/50$ samples,
whereas eligible PC and the D adapter each cover it in $49/50$. PC also equals
the reference in $49/50$; the adapter is broader and never equals it in this
cell. WF produces false assertions in $7/49$ nonempty outputs, compared with
$1/50$ for each comparator. The WF-minus-PC coverage difference is
$-.14\,[-.24,-.06]$. This ordinary-case disadvantage belongs alongside the
joint-effect benefit. Neither broader output nor weaker assumptions alone
ensure better finite-sample recovery.

\subsection{Finite evidence limits both recovery and resolution}\label{sec:finite-results}

Finite CI decisions need not behave like the independences of a single joint
distribution. Substituting them into the order construction can therefore add
or omit graphs relative to exhaustive evaluation of the same decisions.
In the confirmation's XOR-with-descendant condition, with $.10$ flips at both output and descendant
and $N=10{,}000$, practical WF equals the empirical reference family in $36/50$
inputs and covers it in $33/47$ where that
reference is nonempty. The mean fraction of empirical candidates excluded
is $.1196\,[.0576,.1916]$. The discrepancy includes omitted candidates as
well as extra candidates. Exact agreement with this empirical target would still not
establish population coverage.

The Markov confidence family illustrates the information cost of a
broader uncertainty target. In the flip-$.49$ confirmation cell it covers its
Markov reference in $50/50$ samples, but retains a mean of $483.94$ DAGs
out of the four-variable universe of 543, with interval $[449.50,515.56]$.
It does not impose the weaker adjacency or minimality restrictions, so it is
not simply a more accurate version of WF. It preserves uncertainty while
leaving little structural resolution in this regime.

\subsection{Time order narrows explanations under suitable assumptions}\label{sec:temporal-results}

Supplied time order can reduce ambiguity when the model assumptions hold.
In episodes with independent fair baseline variables, an XOR successor and
a noisy-copy successor (Appendix~\ref{app:confirmation-design}), correct
tiers at $N=10{,}000$ reduce mean WF family size from $4.82$ to $1.00$, a paired difference
of $-3.82\,[-4.08,-3.52]$. Equality rises from $43/50$ to $45/50$; both views
have $47/50$ nonempty outputs. The size means include empty families, so a
mean of one does not mean every sample returns one DAG. Nor does a paired
increase in directions establish that those directions arose solely from
orientation propagation: the restrictions can change upstream search.

In the dependent-baseline episode, the second baseline variable instead copies
the first with $.20$ flip probability. With
its true tiers, both the population reference and the returned family are
empty in all 50 samples. Equality is therefore $50/50$, while complete nonempty output
is $0/50$. Its population reference is already empty without tiers: the
generator violates the chosen weaker-adjacency assumption. The all-empty
agreement records incompatibility rather than causal recovery.

For sufficient-history trajectories, two independent binary processes each
flip with probability $.05$ per step. Across 50 trajectory/matched-independent
input pairs at $N=10{,}000$, WF without supplied knowledge has a
trajectory-minus-independent coverage difference of $-.02\,[-.12,.08]$.
This neither establishes equivalence nor validates independent-row inference
for serial data. Full-joint recovery
is unavailable in the omitted-history controls under both sampling schemes.

\subsection{Increasing size limits useful output and completion}\label{sec:scaling}

The scaling results reveal statistical/model and computational limits at
different sizes (Table~\ref{tab:revision-scaling}). Every four-, six- and
eight-variable WF traversal completes, but complete nonempty outputs fall
from 188/200 to 108/200 and 51/200. These are inventories across heterogeneous
conditions, not a pooled accuracy estimate. An empty family is a completed
answer with no retained candidate under the finite procedure; completion
alone does not make it useful.

\input{tables/revision_scaling}

Within the six/eight-variable joint-effect network cells, none of 200 completed
outputs contains the planted graph, although 47 are nonempty. The absence of
a verified population-assumption classification prevents us from separating
finite-test errors from model incompatibility. Ordinary networks vary:
sparse six-variable setting~1 at $N=10{,}000$ contains the planted graph in
$24/25$ outputs, while denser eight-variable setting~2 does so in $1/25$.
A separate six-variable control contains three parallel paths whose blocking
requires a conditioning set of size three. At $N=10{,}000$, WF
finishes all 50 searches, returns 40 nonempty outputs
and includes the planted graph in 39; the depth-two $G^2$ policy
cannot answer the required deeper queries in any of those 50 inputs.
No exact six-variable family-recovery claim follows.

All 400 twelve/twenty-variable WF searches were unfinished when further
computation was discontinued for reporting: 399 had no accepted candidate DAG
and one had a partial result. These outcomes establish neither empty target
families nor the time required for completion.
For ordinary-network setting~1 with twenty variables and $N=2{,}000$, mean
observed standalone CPU is 1,239.97 seconds per input without completion.
The current search can therefore remain unfinished well beyond the
application's one-minute window. Its measured cost does not establish a
complexity lower bound for all algorithms using weaker-faithfulness ideas.

\subsection{Neural controls distinguish retention from application success}\label{sec:neural-results}

The adjacency application gives a conditional operational success
and a negative observed-data result. On the fixed six-unit background,
balanced engineered XOR controls recover the complete parent pair with D
in $19/20$, $20/20$ and $20/20$ samples at 5, 10 and 20\,ms; A--C recover
none. All A--D outputs have zero false edges incident to the synthetic
variables. Matching synthetic parent probabilities to the selected neural
firing proportions makes the pair recoverable by every arm in every sample,
so that regime supplies no incremental protection benefit.

On the 18-unit observed panel, A--D return exactly the same edge
sets within each bin width and no protection decision is accepted. The edge
counts are 94, 141 and 149 at 5, 10 and 20\,ms, respectively. There is no
known neural causal truth with which to label those edges correct, and no
inference of biological absence follows from a threshold that was never
crossed among the evaluated candidates.

\input{tables/revision_application_summary}

%% file: tables/revision_scaling.tex
\begin{table}[tbp]
\centering\small
\caption{Completion and output availability for the WF family estimator. Each row contains 200 confirmation inputs: ordinary and joint-effect networks, two parent settings and $N\in\{2{,}000,10{,}000\}$, with 25 inputs per cell. Counts span heterogeneous conditions and are not pooled accuracy estimates. Unfinished searches were discontinued at a cutoff chosen after computation began, without a common prespecified cumulative runtime limit. An empty family contains no accepted graph; an unfinished search has not established emptiness.}\label{tab:revision-scaling}
\begin{tabular}{rrrrr}
\toprule Variables & \shortstack[r]{Completed /\\total inputs} & \shortstack[r]{Complete\\nonempty} & \shortstack[r]{Complete\\empty} & \shortstack[r]{Unfinished\\at cutoff}\\ \midrule
4 & 200/200 & 188 & 12 & 0 \\
6 & 200/200 & 108 & 92 & 0 \\
8 & 200/200 & 51 & 149 & 0 \\
12 & 0/200 & 0 & 0 & 200 \\
20 & 0/200 & 0 & 0 & 200 \\
\bottomrule
\end{tabular}
\par\smallskip\begin{minipage}{.97\linewidth}\footnotesize Of the 400 unfinished searches, 399 contain no accepted graph and one contains a partial family. Above four variables, population-family and assumption classifications are unavailable, preventing a within-assumption theorem test. Full results for each condition are in Table~\ref{tab:revision-scaling-cells}.\end{minipage}
\end{table}

%% file: tables/revision_application_summary.tex
The family application produced no complete family within its specified
limits. All 740 attempted WF searches reached the 60-second elapsed search
window, remained incomplete and retained no accepted candidate DAG. The
adapter returned one candidate in each of its 730 within-budget outputs,
all still incomplete. Accordingly, local family coverage and compelled-arrow
accuracy are unavailable; zero accepted WF candidates does not establish
an empty target family. These observations describe the bounded implementation
and leave the outcome of longer searches unmeasured.

D returned 585 within-budget selected skeletons, including all three observed
panels. In the descendant controls it retained every planted module edge in
the 49, 47 and 49 evaluable inputs at 5, 10 and 20\,ms, respectively, while
averaging 1.71, 1.81 and 1.59 false synthetic-incident edges. The OR and
independent controls were cleaner (Table~\ref{tab:application-details}).
The observed six-unit graphs contain 21, 27 and 30 adjacencies, again with
no accepted protection decision or biological accuracy target.

The application study was stopped after the active attempt ended, with
2,073 of 4,209 planned method executions recorded. All 2,136 unattempted
executions remain in the analysis. Conditions were evaluated in turn within
each replicate, so every profile/bin is represented. The stopping decision
was made after execution began; the observed subset is not a randomly
selected half-replication design. Appendix~\ref{app:application}
reports availability, cost and all-planned bounds. The family application
contains no B arm, so these D scores do not isolate the incremental benefit
of protection.

%% file: Sections/4_discussion.tex
\section{Discussion}\label{sec:discussion}

Operationalising weaker faithfulness requires more than recognising a
joint-dependence pattern. The pattern must survive the learning procedure,
the output must preserve the explanations permitted by the assumptions,
and the finite decisions must be reliable enough to make those explanations
useful. The two approaches studied here expose these requirements at different
points. The protected adjacency learner supplies a simple, measurable
retention mechanism. The family estimator makes the inferential target more
explicit and sometimes preserves information that the selected skeleton loses.
Both approaches have substantial, identifiable limits.

\subsection{What the operational choices achieve}

The strongest stage result is a mechanism, not just a difference in scores.
Reactivation changes whether a candidate remains available; protection changes
whether a surviving edge is deleted later. In the descendant baseline both
changes are needed within the A--D comparison. This explains why evaluating
only a final protection switch, without matching the preceding search, can
misattribute the improvement. The nuisance-variable comparison adds a second
lesson: useful retention can coexist with selective deletion, so keeping all
candidates is not the only response to a marginal-independence failure.

The practical value depends on the signal. Protection helps strong joint
relationships, can add only false edges in other conditions, and fails below
its fixed threshold. The family estimator's flip-$.45$ result shows that
this threshold is a limitation of the chosen operational rule, not a universal
boundary of weaker-faithfulness inference. Its flip-$.49$ and faithful-chain
results show that removing the threshold does not remove finite-sample error.
A decision not to reject independence can eliminate a genuine alternative;
a completed small family can consequently express unwarranted certainty.

These results support the narrower claim that explicit design choices can
recover useful information in specified regimes. They do not provide a
rule for deciding from a new dataset alone whether protection is worthwhile,
a uniformly superior learner, or a finite-sample graph-coverage guarantee.
Their methodological value is in locating the gains and losses and making
the returned object's meaning explicit.

\subsection{Retention and identification require different outputs}

The XOR example prevents a tempting interpretation of adjacency success.
Recovering the two planted links does not establish that the child is
identified, and the union of all potentially relevant edges is not a graph
family. When several skeletons are compatible, committing to one removes
possibilities that a downstream orientation routine cannot reconstruct.
This is a limitation of what is passed to orientation, even if the routine
itself is correct.

Our exact references implement known population constructions under
explicit assumptions. Their role is to establish what the estimator
is trying to retain and to distinguish errors from genuine ambiguity.
They do not create causal knowledge unavailable under the assumptions.
Changing the assumptions can change the answer substantially: a biased-parent
XOR law excluded by the principal strict-association reading is admitted by
a less restrictive reading. The confirmation's dependent-baseline episode
has no compatible population candidate even before time order is supplied.
These are substantive modelling choices that belong in the interpretation.

A broad Markov confidence family can retain valid explanations while being
too uninformative to settle a particular direction. Conversely, making a
family smaller by imposing different assumptions need not improve recovery.
Accuracy, ambiguity and coverage must therefore be evaluated together rather
than reduced to the number of arrows returned.

\subsection{The demonstrated scope and the remaining research problem}

The positive family results concern small, controlled mechanisms. Completed
six/eight-variable searches already show frequent empty outputs and poor
planted-network recovery in the joint-effect scaling cases; larger searches
then fail to complete. Persistent storage and repeated-state merging address
resource use but do not repair statistical decisions or establish that the
chosen assumptions describe the generating network. A generally useful
family learner would need progress on both inference and computation.
Simply allowing the present searches to run longer is not a demonstrated
solution.

Temporal knowledge has a useful but limited role. It can remove explanations
with backward arrows when the measured variables and history support the
assumed model. It does not establish causal sufficiency, the adequacy of a
single lag or the independence of overlapping observations. Our independent
episodes, serial controls and neural frames therefore have different
interpretations. The neural experiments are tests on one fixed background;
synthetic randomizations and bin widths do not become independent biological
replicates. Useful engineered retention is compatible with no observed
protection benefit and with failure to obtain a complete graph family within
the application's one-minute search window. Longer application runs could
change that availability, but the present evidence does not quantify their
benefit or eventual cost.

The reporting design also limits broad performance conclusions. The main
examples are selected for explanation after analysis, with the complete grid
retained. Cellwise intervals quantify the declared sampling variation rather
than simultaneous superiority. Unfinished searches remain in the inventory,
and observed resource use cannot forecast eventual completion. These limits
allow a clear account of what was attempted without treating the amount of
computation as evidence of practical effectiveness.

These findings give concrete requirements for further operationalisation:
coordinate the stages that act on joint dependence, state which explanations
an output retains, account for uncertainty in finite independence decisions,
and deliver useful answers at the intended scale. The present implementations
provide testable examples of both the gains and the failures that those
requirements address.

%% file: Sections/5_conclusion.tex
\section{Conclusion}\label{sec:conclusion}

This attempt to operationalise weaker faithfulness recovers jointly informative
links by coordinating candidate survival and later protection, and preserves
alternative graph explanations through a distinct family search. The measured
gains come with extra edges, false exclusions and computational failures.
Making these choices and costs explicit turns an abstract relaxation into
a method that can be tested and improved. Reliable inference at larger scales
remains the central unresolved requirement.

%% file: Sections/A_appendix.tex
\section{The population target and its construction}\label{app:theory}

This appendix specifies the principal target, so that ``weaker faithfulness''
does not stand for an unspecified relaxation. Variables have finite alphabets,
and the affirmative exact-law results assume a strictly positive joint law.
Deterministic laws remain labelled boundary diagnostics. The graphical Markov
blanket $\operatorname{MB}_G(x)$ contains the parents and children of $x$ and
the other parents of its children. It differs from an algorithm's current
candidate set.

Write $x\not\!\perp_P u\mid T$ for dependence in $P$. Following the
association definitions of \citet{marx2021weaker}, define
\begin{align}
 a_1(x,u)&\iff\ \forall T\subseteq V\setminus\{x,u\},\quad
                  x\not\!\perp_P u\mid T,\\
 a_2(x;u,v)&\iff\ \forall T\subseteq V\setminus\{x,u,v\},\quad
 \begin{cases}
 x\not\!\perp_P u\mid T\cup\{v\},\\
 x\not\!\perp_P v\mid T\cup\{u\},\\
 u\not\!\perp_P v\mid T\cup\{x\}.
 \end{cases}
\end{align}
Our principal strict two-association predicate is
$s_2(x;u,v)=a_2(x;u,v)\land\neg a_1(x,u)\land\neg a_1(x,v)$.
For every ordered adjacent pair $(x,u)$, the weaker adjacency condition
$A_2(G,P)$ requires a subset $A\subseteq\operatorname{MB}_G(x)$ containing
$u$, of size one or two, such that $A=\{u\}$ satisfies $a_1(x,u)$ or
$A=\{u,v\}$ satisfies $s_2(x;u,v)$. The condition is evaluated at both
endpoints of every edge. A spouse can supply the second member of the
association group; the witness need not consist of two parents.

The principal strictness convention requires failure of both individual
one-associations (the ``neither'' reading). The alternative convention instead
requires $\neg[a_1(x,u)\land a_1(x,v)]$ (``not both''). Neither convention
means marginal independence alone: failure of $a_1$ requires an independence
witness somewhere in the full conditioning-set universe. This explicit
interpretation does not claim to settle authorial intent. It changes the
model domain. For example, let $X\sim\mathrm{Bernoulli}(1/3)$, let $Y$
be an independent fair bit, and set $Z=X\oplus Y\oplus E$ with independent
$.10$ output flips. This one-biased-parent XOR law has no principal candidate
but four under the alternative. Assumption failures remain distinct from
finite-sample failures.

Markov compatibility can be checked through all graphical separations or the
complete joint local-Markov statements. Subgraph minimality is
\begin{equation}
 S(G,P)\iff M(G,P)\ \land\
 \nexists H:\ E(H)\subsetneq E(G),\ M(H,P),
\end{equation}
where $E(G)$ is the directed edge set and $H$ is a DAG on the same variables
$V$, with retained arrows keeping their directions. No adjacency-faithfulness or
knowledge restriction is imposed on $H$ in this test. Checking deletion of
one edge at a time is sufficient: if a proper subgraph is Markov-compatible,
an intermediate graph formed by one deletion is also compatible. This is
not global edge-count minimisation.

For an order $\pi$, let $\operatorname{Pred}_\pi(x)$ be the predecessors of
$x$ and set
\begin{equation}
 \operatorname{Pa}_\pi(x)=\{u\in\operatorname{Pred}_\pi(x):
 x\not\!\perp_P u\mid\operatorname{Pred}_\pi(x)\setminus\{u\}\}.
\end{equation}
Under positivity the resulting graph is the unique inclusion-minimal Markov
graph respecting that order, using the established construction discussed
by \citet[Theorem~3.5]{lam2022greedy}. Every SGS-minimal DAG appears by taking
one of its topological orders; otherwise a smaller compatible graph for that
order would contradict its minimality. Filtering these graphs gives
Equation~\ref{eq:family}. These are exact-law statements. Exhaustive filtering
and order construction were independently compared on the small positive-law
suite; the finite fixtures are implementation checks, not the generality
limit of the mathematical statement or a finite-sample coverage proof.

The principal target does not require the additional orientation-faithfulness
predicate. Its separately recorded sensitivity can exclude candidates admitted
by the principal target. Grouped noncollider evidence is retained as an
existential constraint over actual paths, not turned into a noncollider label
for every component triple. A consensus drawing is never used as a substitute
for the joint candidate constraints.

A Markov equivalence class comprises DAGs with the same skeleton and
unshielded colliders. A completed partially directed acyclic graph (CPDAG)
represents such a full class. Compatible background knowledge can restrict
it to a maximally oriented partially directed acyclic graph (MPDAG)
\citep{bang2023time}. Our class representation is accepted only if its full
extension set equals the retained candidates. Families spanning skeletons
require alternatives; restricted subsets can require residual constraints.
The principal population examples establish ambiguity across classes. The
examined examples that fail within-class MPDAG compression occur under the
alternative strictness convention, so they do not establish that stronger
limitation for the principal population target.

\section{Implementation details for the adjacency learner}\label{app:implementation}

All A--D/R comparisons retain the same column order, grow routine, nominal
CI level and conditioning cap. Grow repeats until no candidate is added;
shrink updates its current blanket during traversal. Reactivation uses the
first separating set encountered and tries a one-variable extension only
when that set is a singleton. The proposed removal is vetoed at the first
extension returning dependence. No claim of traversal invariance is made.

For static and within-successor-slice edges, graph construction requires
reciprocal blanket membership. In the adjacency learner's temporal branch, baseline
target blankets are not learned. A cross-time edge requires the baseline
variable to occur in the successor's candidate blanket; baseline--baseline
edges are excluded, and successor--successor edges are eligible. These are
implementation restrictions, not consequences of temporal order alone and
not the full-joint episode model used in the family reference.

The modified $G^2$ CI statistic sums modified $G^2$ contributions over observed
strata. Zero cells receive a $.5$ pseudocount. The rare-category rule can
merge axes with more than two categories; it leaves binary axes unchanged.
There is no multiple-testing filtering in the primary arms. A separate
baseline-only sensitivity varied the protection threshold to $.1$ or $.5$
bits or applied Benjamini--Hochberg filtering at $q=.05$
\citep{benjamini1995fdr}; these are not the primary confirmation policy.

The directly calculated contrast in Equation~\ref{eq:statistic} is sometimes
called net synergy: a two-source partial information decomposition expresses
it as synergy minus redundancy \citep{williams2010nonnegative}, rather than
the synergy atom of the BROJA decomposition \citep{bertschinger2014quantifying}.
Protection evaluates all pairs in each final target blanket. The retained
identifier has an effect only when its edge survives graph assembly.
With fixed input, traversal, shrink and CI answers, let $E_{\mathrm{keep}}\subseteq E_0$
contain the edges ordinary pruning retains and $E_{\mathrm{prot}}$ contain
the protected identifiers. Protected pruning retains
$E_{\mathrm{keep}}\cup(E_{\mathrm{prot}}\cap E_0)$. The set identity
requires the shared pruning-query pool, which here is independent of earlier
removals. It does not prove a target's graphical blanket or a causal skeleton
is recovered.

\section{Finite evidence, uncertainty and comparators}\label{app:finite}

The selected conditional-randomization policy samples each observed
conditioning-stratum table conditional on its row and column totals and
compares the summed statistic with 199 corresponding null draws. Its
$p$-value is $(1+b)/200$, where $b$ is the number of simulated statistics
at least as large as the observed statistic within the recorded numerical
tie tolerance. Query-specific seeds are fixed by the input seed and
normalised query. A query requiring more than 262,144 table cells is
unavailable. No depth-two restriction is imposed on this selected policy.
The depth-two $G^2$ family-search policy instead uses the modified
likelihood-ratio test and treats deeper-than-two queries as unavailable. Both policies are kept
separate from the protected adjacency learner's own finite decisions.

Three objects must be distinguished from $\mathcal F^*$.
The \emph{empirical family} $\mathcal E$ exhaustively applies graphical
Markov compatibility, the declared association predicates and subgraph
minimality to a complete table of Boolean CI decisions, and imposes the
supplied knowledge $K$. Its four-variable reference
does not assume that the table obeys probability-law identities. The
\emph{practical estimate} uses the order search with finite answers and can
differ from $\mathcal E$ even after completion. Equality to $\mathcal E$
is agreement with this exhaustive finite-decision reference, not statistical coverage of $\mathcal F^*$.

The separate \emph{Markov confidence family} $\mathcal C$ retains every DAG
that does not graphically separate any rejected query from a fixed inventory,
and satisfies the supplied knowledge. For four-variable IID inputs it uses
24 singleton-pair CI queries, 1,999 randomizations and per-query level
$.05/24$. Unavailable and nonrejected queries impose no exclusion. Under
the stated test-validity conditions, the event of no false null rejection
retains every Markov-compatible DAG satisfying the supplied knowledge
simultaneously. This construction does
not impose weaker adjacency faithfulness or minimality. Its broader coverage
and its size therefore answer a different question from WF recovery.
It is ineligible for the overlapping-trajectory diagnostic.

\subsection{Comparator scope}\label{app:comparators}

The D adapter adds no CI tests or collider evidence to the selected skeleton.
For at most four variables, all compatible acyclic orientations are enumerated
and their representation checked. For larger inputs, a search returning one
extension is generally still incomplete, and global family assertions are
withheld. Conditional orientation scores require both a correct D skeleton
and a nonempty compatible reference restricted to that skeleton; the size
of this selected subset must be reported.

PC-stable uses the selected CI policy only in independently classified
faithful four-variable conditions. Its implementation fixes adjacency
sets within each conditioning level. Collider orientation uses the recorded
first separator and can retain order dependence; subsequent representation
checks certify the output class or report failure. Its reported cost includes
exhaustive small-class representation and is not a conventional PC scaling
benchmark.

The checked MD-PC reference implements the published specification independently
for at most four static variables. Singleton-source queries use the
modified $G^2$ policy; joint sources are categorical encodings with ordinal
category merging disabled. Symmetric queries are memoized. An independent
extension check abstains if the output is invalid or its represented class
is empty. An empty graph is one valid candidate, and a nonempty class
excluding the truth can still pass this compatibility screen. Planted truth
is never used for acceptance. The controlled expansion has 18 assumption-admitted
laws and 11 stress laws among its 29 four-variable laws; the confirmation study
uses its own prespecified eligibility. Complete comparator tables and
all abstentions remain in the supplementary tables. Family sizes under MD and WF
are not interchangeable performance targets.

\section{Design details and full displayed-condition tables}\label{app:design}

The controlled expansion changes one axis at a time from three four-variable
baselines: XOR, XOR with descendant and ordinary OR. Parents are independent
fair bits, with output flip probability $.05$; the independent fourth
variable is replaced by a noisy descendant in that mechanism. The 35 conditions
comprise three baselines; eight output-noise changes ($0,.15,.30,.45$ in each
XOR mechanism); six single-parent and six symmetric-parent leakage changes
(with replacement probabilities $.10,.30,.60$); six parent-probability changes
($( .20,.50)$, $(.10,.50)$ and $(.20,.20)$); and six nuisance-dimension changes
(to six/eight variables in each baseline). Leakage replaces the pre-flip
signal with the designated parent, or a fair choice of parent, rather than
resampling the output. Shared innovations pair relevant perturbations.

The 1,400 controlled-expansion inputs yield 14,000 primary static/time-labelled
arm executions; the baseline sensitivities add 480. The distinct initial
120-sample stage experiment is not pooled with that expansion. Figure~\ref{fig:retention}
uses the saved expansion summaries for its matched stage and nuisance panels.
Its descendant and nuisance panels are different mechanisms, not one size
series. The illustrative trace is the saved static descendant input at
$N=10{,}000$, seed 1000. Its D output also retains the false $X-Y$ adjacency;
the trace illustrates recovery and its cost together.
The adjacency application contains 15 observed and 600 engineered
executions. Its synthetic parents are either fair or drawn independently
with probabilities set to the recorded neural proportions; the latter does
not reproduce neural serial dependence.

\subsection{Confirmation mechanisms}\label{app:confirmation-design}

The confirmation's 148 cells and 6,400 input IDs are specified in the supplementary design inventory. The replication count
was reduced before outcomes were generated.
The confirmation study accounts for 62,300 method executions and 75,700 logical
method/view outputs. Each mechanism,
parameter setting, sample size and dimension remains separate in the full
exports. Tables~\ref{tab:revision-family-all-n} and
\ref{tab:revision-scaling-cells} retain every planned sample size for the
main family examples and every scaling cell. Their simple inventory totals
do not supply pooled accuracy or a homogeneous-binomial model.

Table~\ref{tab:confirmation-mechanisms} defines the confirmation examples
used in the prose. All variables are binary. Roots and error bits are
independent except for dependences explicitly shown. A flip probability $e$
means an independent Bernoulli($e$) error added modulo two. Every row denotes
a separate mechanism; confirmation parameters differ from those of the controlled
expansion. The complete inventory preserves the other planned settings.

\input{tables/revision_confirmation_design}

For scaling, setting 1 proposes each earlier node as a parent with probability
$.25$ and caps the parent count at two; setting 2 uses $.65$ and a cap of four.
Each input has its own random topological order. Excess proposed parents are
trimmed uniformly; every nonroot has at least one parent. The root probability
is $.35$. With parent bits $b_1,\ldots,b_k$ in topological order, ordinary
nodes use
\[
 P(X=1\mid b_1,\ldots,b_k)=.10+.80\frac{\sum_{i=1}^k i b_i}{\sum_{i=1}^k i}.
\]
In joint-effect networks, zero-based positions $j\geq2$ with $j\bmod3=2$
have at least two parents and instead use $.10+.80(b_1\oplus b_2)$ if $k=2$,
or $.10+.60(b_1\oplus b_2)+.20\operatorname{mean}(b_3,\ldots,b_k)$ otherwise.
Other nodes retain the ordinary rule. The topology seed belongs to the full
input ID, including $N$; different sample sizes therefore do not reuse one
fixed graph. These definitions do not imply an above-four-variable
weaker-faithfulness classification.

\input{tables/revision_family_all_n}
\input{tables/revision_scaling_cells}

\section{Neural sampling and application endpoint}\label{app:application}

The source is culture 1, recording 2012-02-09-0, in CRCNS ssc-3: one hour
and 166 recorded units. The preprocessing ranks units by total
spike count among those with at least 100 spikes. The 18-unit observed
analysis uses the first 18 selected units. The family application fixes
u69, u64, u58, u78, u68 and u48 at baseline and successor slices. It reuses
the same 12,000 sampled transition-row positions per bin width, without
resampling or changing their stored order.

The 5, 10 and 20\,ms full frames have 719,999, 359,999 and 179,999 transition
rows. The selected sets include 175, 408 and 791 pairs of adjacent source
positions, respectively. Such pairs share a whole bin. These counts are not
effective sample sizes or an inventory of all dependent pairs. Other selected
rows can also remain dependent through the underlying process; shuffling
rows does not establish IID sampling. Positivity, causal sufficiency, stationarity,
adequate history and an acyclic within-bin model remain unverified for the
neural population. No observed edge is labelled biologically correct.

Synthetic modules are independent of the fixed background. Their isolated
references contain three XOR DAGs, five descendant DAGs without knowledge
and one with tiers, one OR collider, or one independent empty graph.
Those are local operational references, not the full reference family of
the combined fifteen/sixteen-variable inputs. Descendant tiers only forbid
backward synthetic-module arcs; they impose neither background directions
nor required adjacencies. There are 1,200 planned engineered inputs and
three observed inputs, with 4,209 method executions and 4,509 logical outputs
because D is reused across knowledge views.

The application specifies limits of 100,000 new WF queries, 500,000 window
states/orders, five million steps and 4,096 active tasks in addition to the
time limits described in the main text. Operating-system monitored
CPU includes child/probe costs and need not equal the worker's own timer.
A saved native result can therefore coexist with a failed within-budget
outcome. The final endpoint and reduced application findings are reported
in Section~\ref{sec:neural-results} and the supplementary application tables.

The application study was stopped after the active attempt ended.
Of 4,209 planned method executions, 2,073 have recorded outcomes and 2,136
were unattempted. The analysis retains all
4,509 logical outputs, including 300 D aliases across knowledge views;
these aliases supply neither independent replication nor additional cost.
All three observed inputs were reached. Every engineered profile/bin has
the first 49 planned replicates represented. The next replicate contains
all five jobs for the 5\,ms descendant input and only D for the 10\,ms
descendant input. The stopping decision was made after execution began;
the planned 100-replicate denominator is retained.

There are 593 recorded D outcomes: 585 within-budget selected skeletons and
eight limited attempts. Seven limited attempts retain native selected-skeleton
outputs, excluded from the primary scores. Among 740 recorded adapter outcomes,
730 return one candidate and remain incomplete; ten are unavailable because
their D dependency was limited. All 740 recorded WF outcomes are incomplete
at the elapsed search-window limit, with zero accepted DAGs. None of the
recorded family outputs is complete or complete empty. Unattempted counts are
610 for D and 763 each for the adapter and WF. Family-recovery and compelled-arrow
scores have no evaluable observations: their all-planned binary bounds remain
$[0,1]$, and corresponding paired differences remain $[-1,1]$.

Recorded application attempts used approximately 93,322 CPU seconds,
including monitored child processes. Reused D outputs contribute no
additional computation. The eight limited D attempts include one storage
interruption and seven CPU-limit outcomes; all remain in the accounting.

\input{tables/revision_application_details}

%% file: tables/revision_confirmation_design.tex
\begingroup\small
\setlength{\tabcolsep}{5pt}
\begin{longtable}{>{\raggedright\arraybackslash}p{.24\textwidth}>{\raggedright\arraybackslash}p{.69\textwidth}}
\caption{Mechanisms behind the reported confirmation examples. For a
two-parent conditional-probability table, entries follow $00,01,10,11$.}
\label{tab:confirmation-mechanisms}\\
\toprule
Condition & Definition\\
\midrule
\endfirsthead
\multicolumn{2}{l}{\textit{Table \thetable\ continued: confirmation mechanisms}}\\
\toprule Condition & Definition\\\midrule
\endhead
\bottomrule\endfoot
Symmetric XOR & Independent fair $X,Y$;
$Z=X\oplus Y\oplus E$, with the stated flip probability for $E$;
independent $W\sim\mathrm{Bernoulli}(1/3)$.\\[5pt]
Asymmetric interaction & Independent $X\sim\mathrm{Bernoulli}(.35)$,
$Y\sim\mathrm{Bernoulli}(.60)$;
$P(Z=1\mid X,Y)=(.10,.90,.80,.20)$;
independent $W\sim\mathrm{Bernoulli}(1/3)$.\\[5pt]
Faithful chain & $X_0\sim\mathrm{Bernoulli}(1/3)$ and
$X_j=X_{j-1}\oplus E_j$ for $j=1,2,3$, with $.10$ flips.\\[5pt]
Ordinary OR & Independent fair $X,Y$;
$Z=(X\mathbin{\mathrm{OR}}Y)\oplus E$, with $.10$ flips;
independent $W\sim\mathrm{Bernoulli}(1/3)$.\\[5pt]
XOR with descendant & Independent fair $X,Y$;
$Z=X\oplus Y\oplus E$, $W=Z\oplus F$, with $.10$ flips for both errors.\\[5pt]
Independent-baseline episode & Baseline $X,Y$ are independent fair bits.
Next slice: $Z=X\oplus Y\oplus E$, $W=Y\oplus F$, with $.10$ flips.
Supplied tiers are $(0,0,1,1)$ for $(X,Y,Z,W)$.\\[5pt]
Dependent-baseline episode & Baseline $X$ is fair and $Y=X\oplus E$
with $.20$ flips. Next slice: $Z=X\oplus Y\oplus F$, $W=X\oplus H$,
with $.10$ flips. The same tiers apply. The principal reference is empty
both with and without these tiers.\\[5pt]
Sufficient-history trajectory & Two independent processes
$X_j[t+1]=X_j[t]\oplus E_{jt}$, $j=0,1$, with $.05$ flips.
Independent fair initial states are exactly stationary. $N+1$ states yield
$N$ overlapping two-slice rows. The matched IID control redraws a stationary
baseline and transition for every row using separate randomness.\\[5pt]
Separator control & $X_0\sim\mathrm{Bernoulli}(1/3)$;
$X_1,X_2,X_3$ copy $X_0$ with respective flip probabilities $.10,.15,.20$.
$P(X_4=1\mid X_1,X_2,X_3)=.05+.30(X_1+X_2+X_3)$;
$X_5$ copies $X_4$ with $.20$ flips.
Separating $X_0$ from $X_4$ requires $\{X_1,X_2,X_3\}$.\\
\end{longtable}
\endgroup

%% file: tables/revision_family_all_n.tex
\small
\begin{longtable}{llrrrrr}
\caption{The four displayed family regimes at all planned sample sizes. Every row has 50 planned inputs. Coverage means retaining the entire population WF reference; equality means returning exactly that reference. Any false counts outputs with at least one false assertion, conditional on complete nonempty output; the actual denominator is shown. Mean family size includes complete empty outputs. All searches in these rows complete. PC is included only in its eligible faithful-chain cells. The underlying saved estimates and cellwise 95\% intervals are supplied in \texttt{family\_metrics\_source.csv}, in the supplement's \texttt{figure\_sources/} directory.}\label{tab:revision-family-all-n}\\
\toprule Condition / $N$ & Method & Covered & Equal & Any false & Nonempty & Mean size\\ \midrule
\endfirsthead
\caption[]{Family comparisons at all planned sample sizes (continued).}\\
\toprule Condition / $N$ & Method & Covered & Equal & Any false & Nonempty & Mean size\\ \midrule
\endhead
XOR: flips .10 / 500 & WF family & 48/50 & 41/50 & 0/48 & 48/50 & 3.18 \\
 & D adapter & 0/50 & 0/50 & 50/50 & 50/50 & 6.00 \\
\addlinespace[2pt]
XOR: flips .10 / 2,000 & WF family & 42/50 & 38/50 & 2/44 & 44/50 & 2.74 \\
 & D adapter & 0/50 & 0/50 & 50/50 & 50/50 & 6.00 \\
\addlinespace[2pt]
XOR: flips .10 / 10,000 & WF family & 42/50 & 38/50 & 1/43 & 43/50 & 2.72 \\
 & D adapter & 0/50 & 0/50 & 50/50 & 50/50 & 6.00 \\
\addlinespace[2pt]
XOR: flips .45 / 500 & WF family & 12/50 & 12/50 & 28/40 & 40/50 & 1.44 \\
 & D adapter & 0/50 & 0/50 & 50/50 & 50/50 & 1.10 \\
\addlinespace[2pt]
XOR: flips .45 / 2,000 & WF family & 35/50 & 31/50 & 5/40 & 40/50 & 2.50 \\
 & D adapter & 0/50 & 0/50 & 50/50 & 50/50 & 1.18 \\
\addlinespace[2pt]
XOR: flips .45 / 10,000 & WF family & 48/50 & 41/50 & 2/50 & 50/50 & 3.26 \\
 & D adapter & 0/50 & 0/50 & 50/50 & 50/50 & 1.04 \\
\addlinespace[2pt]
XOR: flips .49 / 500 & WF family & 0/50 & 0/50 & 49/49 & 49/50 & 1.10 \\
 & D adapter & 0/50 & 0/50 & 50/50 & 50/50 & 1.04 \\
\addlinespace[2pt]
XOR: flips .49 / 2,000 & WF family & 1/50 & 1/50 & 48/49 & 49/50 & 1.08 \\
 & D adapter & 0/50 & 0/50 & 50/50 & 50/50 & 1.06 \\
\addlinespace[2pt]
XOR: flips .49 / 10,000 & WF family & 12/50 & 11/50 & 36/48 & 48/50 & 1.50 \\
 & D adapter & 0/50 & 0/50 & 50/50 & 50/50 & 1.02 \\
\addlinespace[2pt]
Faithful chain / 500 & WF family & 44/50 & 44/50 & 4/48 & 48/50 & 3.82 \\
 & D adapter & 49/50 & 0/50 & 1/50 & 50/50 & 8.12 \\
 & PC & 48/50 & 48/50 & 1/49 & 49/50 & 3.86 \\
\addlinespace[2pt]
Faithful chain / 2,000 & WF family & 42/50 & 42/50 & 7/49 & 49/50 & 3.98 \\
 & D adapter & 49/50 & 0/50 & 1/50 & 50/50 & 8.08 \\
 & PC & 48/50 & 48/50 & 2/50 & 50/50 & 4.16 \\
\addlinespace[2pt]
Faithful chain / 10,000 & WF family & 42/50 & 42/50 & 7/49 & 49/50 & 3.88 \\
 & D adapter & 49/50 & 0/50 & 1/50 & 50/50 & 8.08 \\
 & PC & 49/50 & 49/50 & 1/50 & 50/50 & 4.08 \\
\addlinespace[2pt]
\bottomrule
\end{longtable}
\normalsize

%% file: tables/revision_scaling_cells.tex
\small
\begin{longtable}{lrrrrrrrr}
\caption{Unpooled scaling accounting. Ordinary (O) and joint-effect (J) networks remain separate. Settings 1/2 use parent-selection probabilities .25/.65 and parent caps 2/4, respectively. Each row has 25 planned inputs. Complete nonempty, complete empty and incomplete outputs partition those inputs; planted membership includes complete empty failures. Partial counts incomplete searches with at least one accepted candidate DAG. A dash indicates no complete output. At four variables O is assumption-admitted and J is assumption-stress; above four, the full-joint reference and assumption classification are unavailable.}\label{tab:revision-scaling-cells}\\
\toprule Type & Setting & Variables & $N$ & Nonempty & Empty & Incomplete & Planted & Partial\\ \midrule
\endfirsthead
\caption[]{Unpooled scaling accounting (continued).}\\
\toprule Type & Setting & Variables & $N$ & Nonempty & Empty & Incomplete & Planted & Partial\\ \midrule
\endhead
J & 1 & 4 & 2,000 & 23 & 2 & 0 & 0/25 & 0 \\
J & 1 & 4 & 10,000 & 25 & 0 & 0 & 1/25 & 0 \\
J & 2 & 4 & 2,000 & 18 & 7 & 0 & 0/25 & 0 \\
J & 2 & 4 & 10,000 & 25 & 0 & 0 & 1/25 & 0 \\
O & 1 & 4 & 2,000 & 25 & 0 & 0 & 24/25 & 0 \\
O & 1 & 4 & 10,000 & 23 & 2 & 0 & 22/25 & 0 \\
O & 2 & 4 & 2,000 & 25 & 0 & 0 & 22/25 & 0 \\
O & 2 & 4 & 10,000 & 24 & 1 & 0 & 23/25 & 0 \\
\addlinespace[4pt]
J & 1 & 6 & 2,000 & 4 & 21 & 0 & 0/25 & 0 \\
J & 1 & 6 & 10,000 & 11 & 14 & 0 & 0/25 & 0 \\
J & 2 & 6 & 2,000 & 4 & 21 & 0 & 0/25 & 0 \\
J & 2 & 6 & 10,000 & 14 & 11 & 0 & 0/25 & 0 \\
O & 1 & 6 & 2,000 & 21 & 4 & 0 & 20/25 & 0 \\
O & 1 & 6 & 10,000 & 25 & 0 & 0 & 24/25 & 0 \\
O & 2 & 6 & 2,000 & 12 & 13 & 0 & 3/25 & 0 \\
O & 2 & 6 & 10,000 & 17 & 8 & 0 & 15/25 & 0 \\
\addlinespace[4pt]
J & 1 & 8 & 2,000 & 2 & 23 & 0 & 0/25 & 0 \\
J & 1 & 8 & 10,000 & 8 & 17 & 0 & 0/25 & 0 \\
J & 2 & 8 & 2,000 & 1 & 24 & 0 & 0/25 & 0 \\
J & 2 & 8 & 10,000 & 3 & 22 & 0 & 0/25 & 0 \\
O & 1 & 8 & 2,000 & 15 & 10 & 0 & 14/25 & 0 \\
O & 1 & 8 & 10,000 & 15 & 10 & 0 & 14/25 & 0 \\
O & 2 & 8 & 2,000 & 3 & 22 & 0 & 0/25 & 0 \\
O & 2 & 8 & 10,000 & 4 & 21 & 0 & 1/25 & 0 \\
\addlinespace[4pt]
J & 1 & 12 & 2,000 & 0 & 0 & 25 & -- & 0 \\
J & 1 & 12 & 10,000 & 0 & 0 & 25 & -- & 0 \\
J & 2 & 12 & 2,000 & 0 & 0 & 25 & -- & 0 \\
J & 2 & 12 & 10,000 & 0 & 0 & 25 & -- & 0 \\
O & 1 & 12 & 2,000 & 0 & 0 & 25 & -- & 0 \\
O & 1 & 12 & 10,000 & 0 & 0 & 25 & -- & 1 \\
O & 2 & 12 & 2,000 & 0 & 0 & 25 & -- & 0 \\
O & 2 & 12 & 10,000 & 0 & 0 & 25 & -- & 0 \\
\addlinespace[4pt]
J & 1 & 20 & 2,000 & 0 & 0 & 25 & -- & 0 \\
J & 1 & 20 & 10,000 & 0 & 0 & 25 & -- & 0 \\
J & 2 & 20 & 2,000 & 0 & 0 & 25 & -- & 0 \\
J & 2 & 20 & 10,000 & 0 & 0 & 25 & -- & 0 \\
O & 1 & 20 & 2,000 & 0 & 0 & 25 & -- & 0 \\
O & 1 & 20 & 10,000 & 0 & 0 & 25 & -- & 0 \\
O & 2 & 20 & 2,000 & 0 & 0 & 25 & -- & 0 \\
O & 2 & 20 & 10,000 & 0 & 0 & 25 & -- & 0 \\
\bottomrule
\end{longtable}
\normalsize

%% file: tables/revision_application_details.tex
\begin{table}[htbp]
\centering\small
\caption{D's local adjacency outcomes at the application reporting cutoff. Each profile/bin planned 100 synthetic randomizations on one fixed neural background. D/100 counts within-budget outputs; Exact counts outputs matching all known synthetic-incident adjacencies, including nonedges. Synthetic-incident edges have at least one synthetic endpoint, including links to the neural background. Recall and mean false-edge counts use the same evaluable outputs. Bounds retain all unattempted or limited cases and refer to all 100 planned inputs. Independent controls have no planted edges, so recall is undefined. These are local operational scores, without biological accuracy or an incremental D-versus-B comparison.}\label{tab:application-details}
\begin{tabular}{lrrrrrr}
\toprule
Profile & Bin (ms) & D/100 & Exact & Recall & False edges & Exact bounds\\
\midrule
XOR + descendant & 5 & 49/100 & 0/49 & 1.000 & 1.714 & $[0.00,0.51]$\\
 & 10 & 47/100 & 0/47 & 1.000 & 1.809 & $[0.00,0.53]$\\
 & 20 & 49/100 & 2/49 & 1.000 & 1.592 & $[0.02,0.53]$\\
\addlinespace[4pt]
Symmetric XOR & 5 & 48/100 & 0/48 & 0.958 & 1.000 & $[0.00,0.52]$\\
 & 10 & 49/100 & 0/49 & 0.939 & 1.020 & $[0.00,0.51]$\\
 & 20 & 49/100 & 0/49 & 0.990 & 0.980 & $[0.00,0.51]$\\
\addlinespace[4pt]
OR collider & 5 & 49/100 & 48/49 & 1.000 & 0.020 & $[0.48,0.99]$\\
 & 10 & 49/100 & 48/49 & 1.000 & 0.020 & $[0.48,0.99]$\\
 & 20 & 49/100 & 47/49 & 1.000 & 0.041 & $[0.47,0.98]$\\
\addlinespace[4pt]
Independent & 5 & 48/100 & 44/48 & -- & 0.083 & $[0.44,0.96]$\\
 & 10 & 47/100 & 39/47 & -- & 0.170 & $[0.39,0.92]$\\
 & 20 & 49/100 & 44/49 & -- & 0.102 & $[0.44,0.95]$\\
\bottomrule
\end{tabular}
\end{table}